\documentclass[aps,showpacs,preprintnumbers,amsmath,amssymb, 
twocolumn, tightenlines,
]{revtex4}

\usepackage[dvips]{graphicx}
\usepackage{bm}

\begin{document}

\bibliographystyle{apsrev} 

\title {Light bound states of heavy fermions}

\author{M.Yu.Kuchiev} \email[Email:]{kmy@phys.unsw.edu.au}
\author{V.V.Flambaum} \email[Email:]{flambaum@phys.unsw.edu.au}

\affiliation{School of Physics, University of New South Wales, Sydney
  2052, Australia}

\author{E. Shuryak} \email[Email:]{shuryak@tonic.physics.sunysb.edu}

\affiliation{Department of Physics, State University of New York,
 Stony Brook, NY 11794, USA}

    \date{\today}

    \begin{abstract} 
In the Standard Model, a group of heavy fermions, e.g. top quarks,
can collectively strongly affect the Higgs field and create relatively
long-lived bound states. 
If there exist new generations of fermions with masses beyond 1 TeV,
strong binding of several of them can make them lighter than even a
single heavy fermion. Using the mean field approximation we find
 multi-fermion states with masses $M \sim 5v\sqrt{N}\approx 1.2\,\sqrt
 N\,$  TeV, with $N=2,3\dots$ being the total number of  heavy
 fermions bound together, and $v=246$ GeV the Higgs VEV. The 
 experimental search for
 multi-fermions within the range of energies $2-3$ Tev would either
 discover them, or suggest absence of new Standard Model
 fermions with larger masses. Possible implications related to
multi-top states and baryonic asymmetry of the Universe are discussed.

    \end{abstract}


    \maketitle


The possibility that there may exist new generations of fermions, leptons and quarks, inspires search for quarks of the forth generation, see Ref. \cite{Amsler:2008zz} and references therein. Since masses of known fermions cover a wide range of energies, one can/should anticipate that there exist super-heavy fermions, with masses $m$  beyond 1 Tev. We show that if they exist, their strong interaction with the Higgs field produces bound states for several heavy fermions (called multi-fermions below).  The multi-fermion masses are relatively light, starting from 2-3 Tev for arbitrary large mass of heavy fermions. 
There exists therefore an opportunity for detecting super-heavy fermions via much lighter multi-fermions. This effect stems from the fact that fermion masses are proportional to the Higgs field. We will see that the Higgs field is strongly suppressed inside the multi-fermion which greatly reduces its mass.

Implications related to strong interaction of fermions with bosons, in particular with the Higgs, which can modify its VEV, have been discussed in
Refs. \cite{%
Vinciarelli:1972zp,%
PhysRevD.9.2291,%
Chodos:1974je,%
Creutz:1974bw,%
Bardeen:1974wr,%
Giles:1975gy,%
Huang:1975ih,%
PhysRevD.15.1694,%
PhysRevD.25.1951,%
PhysRevLett.53.2203,%
PhysRevD.32.1816,%
MacKenzie:1991xg,%
Macpherson:1993rf}. 
There were developed the MIT-bag \cite{Chodos:1974je}, SLAC-bag  \cite{Bardeen:1974wr}, and bag-version of \cite{Creutz:1974bw}. The Freiburg-Lee model \cite{PhysRevD.15.1694} allows one to consider the MIT and SLAC models as its limiting cases.
Our starting point, as well as the important soliton-type object, which appears in our analysis, is close to that discussed in \cite{PhysRevD.15.1694}, as well as in a number of mentioned above works. An advantage of the present work is the analytical solution, which we first derive for a large mass $m > v$ and 
then verify its applicability for a wide range of fermion masses and numbers of heavy fermions using numerical analysis. These results provide clear physical picture of the problem, and give reliable estimates that may be used in the future for search of multi-fermions. 
Recently the idea that the quark can modify Higgs in its vicinity was tested for a single top and topponium
\cite{
MacKenzie:1991xg,%
Macpherson:1993rf} but only minor effects were
found. Frogatt et al \cite{Froggatt:2008ns} 
provided simple Hydrogen-atom-like estimates 
for a system of $N=12$ top quarks (6 top and 6 anti-top for the 
lowest $S_{1/2}$ orbital), which suggested that its binding energy can be large. Unfortunately, our more accurate mean field  calculations \cite{Kuchiev:2008fd}
indicated that the binding is nonzero, though small, only for massless Higgs, while account of the Higgs mass, which is limited by the current experimental bounds, makes this 12 top-antitop system unbound.

Using the Standard Model, consider $N$ heavy fermions that interact with the Higgs field. Take the conventional unitary gauge, in which the Higgs field $\Phi$ is represented by the real field $ \xi$
\begin{equation}
\Phi\,=\,\frac{v}{\sqrt{2}}\left( \begin{array}{c} 0 \\ \xi \end{array}\right)~.	
\label{Phi-xi}	
\end{equation}
Here $v$ is the VEV, which is achieved when $\xi=1$. The Lagrangian for the system of the Higgs and fermion fields $\xi$ and $\psi$ reads ($\hbar=c=1$)
\begin{align}
\!{\cal L}=
\frac{v^2}{2}\Big(\partial^\mu {\xi} \partial_\mu {\xi}\!-\!\frac{m_{\mathrm H}^2}{4}(\xi^2\!-\!1)^2\Big)\!+
\bar \psi( i\gamma^\mu \partial_\mu \!-m \xi ) \psi\,.
\label{L}
\end{align}
The important feature of the problem is
that either (i) the fermion mass $m$, or (ii) their number $N$
or (iii) both are presumed large.
In all the cases the impact on
  the Higgs field is strong, shifting its value away from the vacuum
VEV. We describe them here using the mean field approximation,
though we are aware of effects due to (weak and strong) gauge field
forces as well as corrections 
beyond the mean Higgs field,  which include
many-body, recoil, and relativistic 
retardation/radiative effects, 
which would be addressed elsewhere \cite{largeN}.
The magnitude of these corrections is different for cases (i)-(iii) because of
their different nature and depends on various small
parameters such as $v/m$, $m_{\mathrm H}/m$, and $1/N$. 
The case of large $N>6$ (many top quarks)
needs, in particular,  an extensive discussion,
as many levels are occupied subsequently.

Adopting this approach we replace the fields in Eq.(\ref{L}) by stationary wave functions $\xi$ for the Higgs and $\psi$ for the single-particle wave functions of fermions. Searching for the spherically symmetrical solution, take $\xi=\xi(r)$ and assume that all fermions occupy the same shell with total and angular momenta $j$ and $l$, which is described by the same large $F(r)$ and small $G(r)$ components of the Dirac spinor. Using Eq.(\ref{L}) one writes the Hamiltonian $H$ of the system
\begin{align}
H\,=\,&\int_0^\infty\,
\Big[ \,\,\frac{v^2}{2}\,\Big(\,\xi'^2+\frac{1}{4}\,m_{\mathrm H}^2\,(\xi^2-1)^2
\,\Big)\,(4 \pi\,r^2) 
\label{HFG}
\\
& + N\,\Big(\,2\big(F'+\frac{\varkappa}{r} F\,\big)\,G+m \xi\, (\,F^2-\,G^2\,)\,\Big)\,\Big]\,dr~.
\nonumber
\end{align}
Here $\varkappa=\pm\,(j+1/2)$ for $l=j\,\pm 1/2$, and normalization $\int_0^\infty(F^2+G^2)\,dr=1$ is taken. 
From Eq.(\ref{HFG})  one derives the 
mean-field equations for $\xi,F,G$
\begin{align}
\xi''+\frac{2}{r}\,\xi'&+\frac{m_{\mathrm H}^2}{2}\,\,\xi\,(1-\xi^2)\,=\,\frac{(N-1)m}{4\pi \,v^2}\,\frac{F^2-G^2}{r^2}~,
\label{xi}
\\
&(\varepsilon-m\,\xi)\,F~\,=\,-G'\,+\,(\varkappa/r)\,G~,
\label{F}
\\
&(\varepsilon+m\,\xi)\,G~\,=\,\,~~F'\,+\,(\varkappa/r)\,F~,
\label{G}
\end{align}
where the eigenvalue $\varepsilon$ is presumed positive \cite{minus}. 

It is interesting that the system considered reacts to the presence of the large fermion mass in such a way as to eradicate its influence on its physical parameters. This phenomenon employs three steps. Firstly, the Higgs develops a node on a sphere of radius $r_0$, $\xi(r_0)=0$, so that it is positive outside the sphere, taking at large distances the classical value $\xi=1$, but is negative inside. Secondly, the fermions use this node of the Higgs as an opportunity to be localized in its vicinity, on the surface of the sphere, so that their density inside the sphere is low (hollow sphere). As a result the term $m\xi$ in Eqs.(\ref{F}),(\ref{G}) is suppressed. Thirdly, the fermion wave function is tuned to satisfy $F^2(r)\simeq G^2(r)$, which eliminates the term $\sim m$ from Eq.(\ref{xi}). Thus, the described configuration of fields suppresses $m$ everywhere in Eqs.(\ref{xi})-(\ref{G}).

To justify this physical picture analytically consider the large fermion mass, $m\gg v$. Assume that some smooth function $\xi(r)$, which has a node at $r_0$, is given. Search for the solution of the Dirac equation in the form
\begin{align}
&F(r)\,=\,A(r) \,\exp(-S(r)\,)~,
\label{FA}
\\
&G(r)\,=\,B(r) \, \exp(-S(r)\,)~,
\label{GA}
\\
S(r)\,&=\,m\,\int_{r_0}^r\xi(r')dr'\,\simeq\,\frac{1}{2}\,m \,\xi'(r_0)\,(r-r_0)^2~.
\label{S}
\end{align}
The large mass $m$ here favors localization of $F(r)$ and $G(r)$ in the vicinity of $r_0$, which justifies the last identity in (\ref{S}). From (\ref{F})-(\ref{S}) one finds equations on $A,B$.
\begin{figure}
\centering
\includegraphics[height=4.5 cm,keepaspectratio = true, 
]{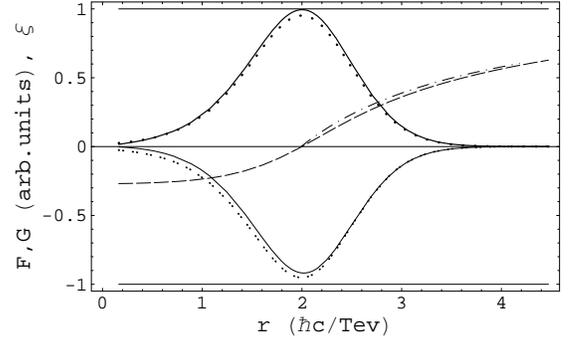}
\caption{ \label{ras} Thick and dotted lines - large component $F(r)$ of the Dirac spinor numerically and analytically, thin and double-dotted lines - same for small component $G(r)$, slashed and slash-dotted lines - same for Higgs $\xi(r)$; numerical data - from solution  of Eqs.(\ref{xi})-(\ref{G}), analytical data for fermions from Eqs.(\ref{FA})-(\ref{S}),(\ref{ABC}), analytical $\xi(r)$ is found from the minimization of ${\cal H}$ in (\ref{h}) as explained in the text.
}
\end{figure}
To verify that $A$ and $B$ are smooth functions of $r$ in the vicinity of $r=r_0$ (which guarantees that fermions are localized near $r=r_0$) we expand the coefficient functions 
$\varkappa/r$ and $\xi(r)$ in (\ref{F})-(\ref{G}) in powers of $r-r_0$, taking first the lowest-order approximation $\varkappa/r\approx \varkappa/r_0$, $\xi(r)\approx \xi'(r_0)(r-r_0)$. Then one finds the eigenvalue
\begin{equation}
	\varepsilon\,=\,-\varkappa/r_0\,=\,(j+1/2)/r_0\,>\,0~,
	\label{eps}
\end{equation}
which does not grow with increase of $m$, and is positive provided $\varkappa<0$.
The corresponding functions $A(r),B(r)$ turn constants, $A(r)=-B(r)=const$. 
Normalizing them using Eqs.(\ref{FA})-(\ref{S}) we find
\begin{equation}
A\,=\,-\,B\,=\,[\,m\xi'(r_0)/4\pi\,]^{1/4}~.
	\label{ABC}
\end{equation}
Using higher-order expansion for $\varkappa/r$ and $\xi(r)$ in powers of $r-r_0$ we verified that Eqs. (\ref{eps}) and (\ref{ABC}) remain valid for $m\gg v$. Thus, Eqs.(\ref{FA})-(\ref{ABC}) give analytical solution   of the Dirac equations (\ref{F}),(\ref{G}) when $m > v$. 
Fig. \ref{ras} shows good agreement of the found analytical wave functions with results of numerical calculations discussed in more detail below.

Using the found eigenvalue (\ref{eps}) of the Dirac equation we can simplify the Hamiltonian in Eq.(\ref{HFG}), where the term in the second line equals $N\varepsilon= N|\varkappa\,|/r_0$. It is convenient at this stage to scale distances by the Higgs mass, $r\rightarrow x= m_{\mathrm H}\,r$,
and present $H$ in the form
\begin{align}
&H\,=\,\frac{v^2}{m_{\mathrm H}}\,{\cal H}~, 
\label{Hh}
\\
&{\cal H}\,=\,4 \pi\,\int_{x_0}^\infty\,
\Big(\,\frac{\xi'^2}{2}+\frac{1}{4}(\xi^2-1)^2\,\Big)\,x^2
 \,dx+\frac{z}{x_0}~.
\label{h}
\end{align} 
Here $\xi=\xi(x)$, $\xi'=d\xi/dx$, and $x_0=m_{\mathrm H}\,r_0>0$ is the node, $\xi(x_0)=0$. The second term in Eq.(\ref{h}), in which 
\begin{equation}
z\,=\,-N\,\varkappa\,\,m_{\mathrm H}^2/v^2\,>\,0~, 	
	\label{z}
\end{equation}
reproduces the second term from Eq.(\ref{HFG}), which equals $N\varepsilon$ (as was mentioned).  The integration in Eq.(\ref{h}) neglects the contribution of distances $x<x_0$. This approximation proves convenient and at the same time accurate since  $\xi$ is small and smooth for $x<x_0$, while the factor $x^2$ in the integrand produces strong suppression compared to the outer region $x>x_0$ \cite{smallx0}.
\begin{figure}[t]
\centering
\includegraphics[height=4.5 cm,keepaspectratio = true, 
]{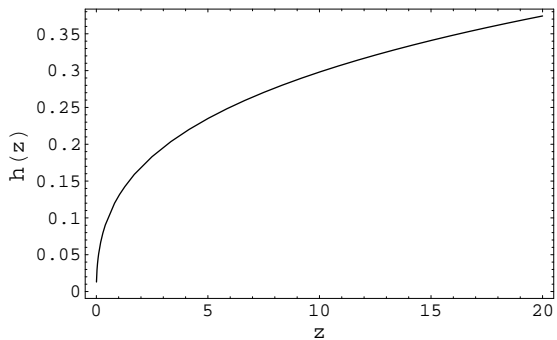}
\caption{
 \label{dva} The etalon function $h(z)$ presents the energy of a multi-fermion in Eq.(\ref{Hh}) via the dimensionless quantities, ${\cal H}(z)$  and $z$ from Eqs.(\ref{h(z)}),(\ref{z}). 
}\end{figure}
\noindent

The energy $H$ of the system is presented in Eqs.(\ref{Hh}),(\ref{h}) as a functional of $\xi(x)$.  Finding its minimum we derive the final answer to the multi-fermion problem in terms of simple, dimensionless quantities ${\cal H}$ and $z$. To find this minimum we minimized firstly ${\cal H}$ over $\xi(x)$ keeping the node $x_0$ fixed. This required solution of the boundary problem, $\xi''+(2/x)\,\xi'+\xi\,(1-\xi^2)/2=0$, $\xi(x_0)=0$, $\xi(\infty)=1$. The result found for ${\cal H}$ was consequently minimized over $x_0$. The outcome is the function ${\cal H}={\cal H}(z)$, which we calculated numerically, depicting result in Fig. \ref{dva} in terms of the {\it etalon} function $h(z)$, which is related to ${\cal H}(z)$ via
\begin{equation}
{\cal H}(z)\,=\,2\,(2\pi z)^{1/2}\,(\,1+h(z)\,)~.
	\label{h(z)}
\end{equation}
The first factor  here equals the lowest term of the expansion of ${\cal H}(z)$ in powers of $z^{1/2}$, ${\cal H}(z)\rightarrow 2\,(2\pi z)^{1/2}+O(z)$ 
(with $x_0 \rightarrow [z/(2\pi)]^{1/2}$), indicating that $h(0)=0$. 
It can be shown that  $h(z)$ reveals the asymptotic behavior  $h(z)\sim z^{1/6}$ for $z\rightarrow \infty$, 
but within a wide range of $z$ the function $h(z)$ remains small, $h(z)\ll 1$, see Fig.\ref{dva}. 
From Eqs.(\ref{Hh})-(\ref{h(z)}) we derive the analytic expression  for the total mass $M_a$ of the system of $N$ heavy fermions, which compose the multi-fermion
\begin{equation}
	M_a=2\,(2\pi N |\varkappa\,|)^{1/2}v\,\big[ \,1+h(N|\varkappa\,|m_{\mathrm H}^2/v^2\,)\,\big]~.
	\label{Ma}
\end{equation}
Here the term $\sim h(z)$ remains small for a wide range of values of $N,j$ and $m_{\mathrm H}$, see Fig.\ref{dva}. Consequently, the scale of multi-fermion masses is defined mainly by the Higgs vacuum expectation value $v=246$ Gev, which gives
$M_a \approx 5.01 \,(N |\varkappa\,|\,)^{1/2}v= 1.23\,(N |\varkappa\,|)^{1/2}$ Tev. 
\begin{figure}[b]
\centering
\includegraphics[height=4.5 cm,keepaspectratio = true, 
]{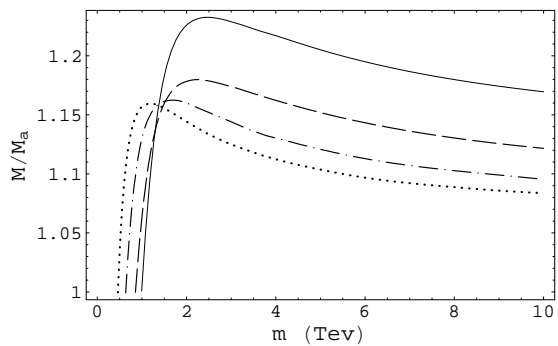}
\caption{ \label{tri} Masses of multi-fermions $M$ vs  heavy fermion mass $m$, $M$ is scaled by $M_a$ from Eq.(\ref{Ma}). Solid, dashed, dashed-dotted, and dotted lines - $N=2,3,6,$ and $12$ respectively, $j=1/2$, $l=0$, $m_H=100$ Gev.
}
\end{figure}

Consider numerical solution of the self-consistent mean field Eqs.(\ref{xi})-(\ref{G}).   Proper formulation of this problem includes boundary conditions, which for fermions have the conventional form, whereas for the Higgs field they read
\begin{align}
\xi'(0)\,=\,0~,\quad \quad \xi(\infty)\,=\,1~.
\label{xi0}
\end{align}
The first one suppresses the singularity in the term $(2/r) \xi'$ in Eq.(\ref{xi}) at $r=0$. In linear equations this singularity is harmless, it is eliminated by the conventional scaling of the function by $\sim 1/r$ factor. For nonlinear equations this, and others tricks do not work, the singularity persists making $\xi(0)$ singular. The first condition in (\ref{xi0}) removes this nuisance, allowing $\xi(0)$ to be finite, as it should be from the physical point of view.

Our numerical calculations  were performed taking $m_\mathrm{H}=100$ Gev for $j=1/2$, $l=0$, $N=2,3,6$ and $12$ \cite{annihilation}. 
Fig. \ref{ras} shows wave functions 
for $N=3$ and $m=10$ Tev. The good agreement of analytical and numerical results supports the fact that fermions in the multi-fermion occupy mostly the surface of the sphere, where the Higgs field develops a node being positive outside the sphere and negative inside. Fig. \ref{tri} shows the calculated mass $M$ of  multi-fermions for different $N$ and $m$. It is presented as a ratio $M/M_a$, where $M_a$ is the analytical result (\ref{Ma}). In the region of small $m$ the multi-fermion mass $M$ falls, and for sufficiently small $m$ the
 bound state disappears. With increase of $m$ the mass $M$ shows a smooth maximum, after which it decreases, revealing a tendency to converge to the analytical result (\ref{Ma}).
Fig. \ref{tri} shows that for a wide area of variation of $m$ and $N$ the masses of  multi-fermions are close to the analytical predictions of Eq.(\ref{Ma}).

Remarkably, the multi-fermion masses found in Eq.(\ref{Ma}) do not grow with an increase of the heavy fermion mass $m$. For $j=1/2$ Eq.(\ref{Ma}) predicts masses 
$ 1.7\times K_2 $ Tev where $1.06<K_2<1.17$ and $2.1 \times K_3$ Tev where
$1.07<K_3<1.21$ for multi-fermions constructed from two and three heavy fermions respectively; the coefficients $K_2$ and $K_3$ estimate corrections produced by the second term in the brackets in Eq.(\ref{Ma}), when the Higgs mass spans the interval 100-300 Gev. Fig. \ref{tri} shows that 
if the fermion mass $m$ is not very large, then the masses of multi-fermions are slightly larger, by up to 25\%.  

For possible experimental applications it is interesting that for a small  number of heavy fermions, $N=2,3$, the masses of multi-fermions belong to the interval of energies of 2-3 Tev. This is in contrast to heavy fermions themselves, whose masses may be larger. 
This fact provides an opportunity for hunting for super heavy fermions indirectly, via discovery of relatively light multi-fermions. A closely connected topic is related to bubbles made of a large number of top-quarks, or other heavy fermions (or bosons like $Z,W$). Note  that the derivative $dM/dN$ for the multi-fermion decreases with $N$. This means that the lifetime
for the weak decay of a single fermion inside the bubble increases fast with $N$ because
of decrease of the released energy $M(N)-M(N-1)$. This  makes ``magic'' multi-fermions
with complete  or nearly complete $1s,2p_{3/2},...$ shells  relatively
long-lived states.   Note that for a very large $N$ the summation over
occupied shells can be carried out analytically and may be expressed via 
the following redefinition of the parameter $z$ in Eq.(\ref{h}), 
$z \approx (2/3) N^{3/2}N_c^{-1/2}\,m_{\mathrm H}^2/v^2$ where $N_c$ is the number of
different types of fermions involved (e.g. 3 colors for quarks).
Then the mass is given by Eqs. (\ref{Hh}),(\ref{h(z)}).

Another intriguing implication is related to the baryonic asymmetry
created during the Big Bang. Although in principle all Sakharov
conditions are satisfied in the Standard Model,
numerically they are extremely restrictive. Nevertheless, 
as near the electroweak phase transition the bubbles with different
values of the Higgs field are created, the top 
 (and possibly heavier) quarks get  bound to the zero-Higgs surfaces,
 leading to long-lived states and large local deviations from thermal
 equilibrium. This effect may substitute for the ``bubble walls''
 (much discussed in literature when the 1st order phase transition
 was still an option). The CP violation and baryon number asymmetry in
 the decay of these states presents an appealing problem worth studying. 

ES would like to thank Holger Nielson
for inspiring talk and subsequent discussion
which sparked our interest in the subject.
The work of VF and MK is supported by the Australian Research Council
and that of ES by US-DOE grants DE-FG02-88ER40388 and
DE-FG03-97ER4014.


\begin{thebibliography}{10}
\bibitem{minus}
There are known fermion states with negative energy, as the ones produced by
  superheavy nuclei for electrons, but we do not encounter such states here.
\bibitem{smallx0}
This is valid provided $x_0$ is not extremely large, say $x_0<50$, which is
 the case for all examples discussed below.
\bibitem{annihilation}
If fermions and antifermions are present in one bound state, their virtual
  annihilation makes their attraction stronger. For simplicity, we
  neglect this effect here noting only that it can reduce masses of
  multi-fermions.
\end{thebibliography}
\end{document}